\documentclass[12pt]{article}
\begin{document}

\newcommand{\ind}{\hspace{0.3in}}
\newcommand{\noind}{\hspace{-0.3in}}
\newcommand\lsim{\mathrel{\rlap{\lower4pt\hbox{\hskip1pt$\sim$}}
    \raise1pt\hbox{$<$}}}
\newcommand\gsim{\mathrel{\rlap{\lower4pt\hbox{\hskip1pt$\sim$}}
    \raise1pt\hbox{$>$}}}

\begin{flushright}
SHEP 97/11\\
hep-ph/9707317\\
July 1997\\
\end{flushright}

\vspace{.4in} 

\begin{center}
\Large\textbf {Challenges for Inverted Hybrid Inflation}

\vspace{0.2in}

\normalsize
S. F. King and J. Sanderson

\emph{Department of Physics and Astronomy, University of Southampton,\\
Southampton, SO17 1BJ, U.K.}

\end{center}

\vspace{0.1in}

\footnotesize

\begin{center}
\emph{ABSTRACT}
\end{center}

Inverted hybrid inflation (in which the inflaton field slowly rolls away
from the origin, giving a spectral index $n\lsim 1$) is
an appealing variant of the more usually studied hybrid
model. Analysing the model alongside the ordinary hybrid case, we show
that, in order to provide the correct density perturbations
consistent with the COBE measurements, the 
dimensionless coupling constants of the
inverted hybrid potential must be very small indeed.
For example, the quartic coupling in a typical such potential is found to be
$\lsim 10^{-12}$.
A supersymmetric model of inverted hybrid inflation, which
does not involve the troublesome quartic coupling is found to lead to 
a potential which is unbounded from below.

\vspace{.5in}

\normalsize 

\newpage

\section{Introduction}

\ind The hybrid inflation scenario proposed by Linde
\cite{ref1,ref2} is probably \emph{the} most attractive model of
inflation, at present. Not only does it remove any necessity for
fine-tuning, but, after many years, it reconnects inflation with
particle physics, as in the original models of inflation developed in
the early 1980s \cite{ref3}. In particular, it is potentially well suited to
realizations in supersymmetry (SUSY).

The relevance of hybrid inflation to SUSY models has been widely
discussed \cite{ref4, ref4a}. Inflation generally requires a flat
potential in order to satisfy the slow roll conditions. More exactly,
one requires flat directions in field space. If one wishes to avoid
fine tuning at each order in perturbation theory, there are two strategies
to achieve flat directions: Goldstone bosons (which roll around the
flat bottom of the wine bottle potential), or flat directions in 
SUSY theories. The importance of SUSY is that flat directions are not
spoiled by radiative corrections due to the non-renormalisation theorem.
In the context of SUSY the renormalisable scalar potential is a 
sum of squares of F-terms and D-terms, which may both vanish identically
along certain directions in field space called the flat directions.
The massless chiral superfields which parametrise the flat directions
are known as the moduli fields. In fact in realistic theories, the
flat directions are not exactly flat since they are lifted by
two effects: SUSY-breaking effects 
(such as soft masses for the moduli fields), 
and effects due to non-renormalisable terms in the superpotential.
For example, in a particular class of ``supernatural'' theories
of hybrid inflation \cite{ref4a}, 
the renormalisable superpotential is postulated to be zero, and 
the moduli fields are assumed to have zero D-terms, with the potential
for the  moduli fields being provided by SUSY-breaking effects. This
occurs via the
Kahler potential in the context of supergravity theories based
on hidden sector SUSY-breaking at an intermediate scale, $M_I\sim 10^{11}$ GeV.
\footnote{There are well known difficulties with such models based on 
supergravity due to the flatness parameter $(M_P^2V'')/(8\pi V)$ 
being of order unity as a result of the non-renormalisable 
supergravity contributions, but we
shall not address this problem here.} 
However this is just one class of model, and more generally 
it is not necessary to have a vanishing superpotential, 
nor is it necessary to identify all the fields in the superpotential
as moduli fields of flat directions. It is sufficient in hybrid inflation
for the potential to have a flat direction along the ``inflaton'' direction
in the case when the other field takes a zero value. To understand
this statement it is necessary to explain a little more about
hybrid inflation in general terms.

In hybrid inflation more than one scalar field is relevant to
inflation. A scalar field $\psi$ whose eventual vacuum expectation value 
(VEV) is, say, of order $10^{11}$ GeV is initially prevented from
attaining this VEV by a second scalar field $\phi$ which 
takes rather large values
initially (but well below the Planck scale $M_P\approx 1.2\times 10^{19}$ GeV)
as it rolls towards its eventual VEV
at the origin. While the ``inflaton'' field $\phi$ 
takes values larger than some critical value $\phi_c$, the field $\psi$ is
pinned at the false vacuum $\psi = 0$ and inflation takes place.
When the inflaton reaches its critical value
$\phi = \phi_c$, a cascade takes place of the fields to their
eventual VEVs $\langle\psi\rangle\sim 10^{11}$ GeV and $\langle\phi\rangle =0$,
and inflation is abruptly ended.
Consequently this scenario is able to avoid the
standard naturalness problems of single-field models. Not surprisingly
then, hybrid inflation has received a good deal of attention recently and
many different versions have resulted \cite{ref5,ref6,ref7,ref8}. 
Generally, hybrid models give a blue tilted spectrum
of adiabatic density perturbations, corresponding to a spectral index
$n\gsim 1$. In the  context of SUSY theories, it is not necessary to identify
both $\psi$ and $\phi$ as the scalar components of moduli fields.
All that is required is that when $\psi=0$ the potential is flat
in the $\phi$ direction, with the flatness lifted by a soft SUSY breaking
mass $m^2|\phi |^2$, where $m \sim 1$ TeV. In other words only $\psi =0$,
$\phi \neq 0$ must be a SUSY flat direction, with $\phi$ being the scalar
component of a moduli field.

Given the success of hybrid inflation, it was subsequently suggested
that the hybrid mechanism could be
adapted to create an ``inverted'' model in which the inflaton
field $\phi$ has a negative mass squared and
rolls away from the origin, predicting a spectral index which
can be significantly below 1 in contrast to virtually all other hybrid
models. In this ``inverted hybrid inflation'' \cite{ref5,ref6}
the field $\phi$ is supposed to obtain a non-zero VEV eventually,
and this is typically achieved by adding to the potential a $\phi^4$ term.
The purpose of the present paper is to point out that this model is not viable
unless all its dimensionless couplings are extremely small.
Already at a qualitative level, the inverted hybrid scenario looks
mildly uncomfortable since the quartic inflaton coupling does not fit
in with the notion of flat directions outlined above.
Even given this assumption, it is almost obvious without doing any calculation
that the inverted hybrid scenario cannot be viable if all the
dimensionless couplings are of order unity. The argument runs as follows.
Let us assume that all dimensionless couplings are of order unity. Then
if the mass of the $\phi$ field is of order 1 TeV, the  natural scale
for its VEV (in the limit that the $\psi$ field is set to zero)
must also be about 1 TeV. However we know that in inverted hybrid inflation
as in hybrid inflation, for a $\phi$ mass of order 1 TeV, the
$\psi$ mass and VEV must be around $10^{11}$ GeV (still assuming
all dimensionless couplings of order unity), and so the critical
value of the $\phi$ field must also be around $10^{11}$ GeV.
Now we can see the problem, - the $\phi$ field which is rolling from the
origin towards its critical value must necessarily pass through
its natural minimum at 1 TeV, and so it will get hung up there and never reach
its critical value. The only way round this problem is by fine tuning the
dimensionless couplings so that the critical value of $\phi$ is below the
$\phi $ VEV (in the limit that $\psi=0$) and this leads to the stated
conclusion that in inverted hybrid inflation the
dimensionless couplings are necessarily extremely small.

The remainder of the paper consists of fleshing out the above argument.
In sections 2 and 3 we shall focus on a
definite form of the potential which enables easy comparison
between the hybrid and inverted hybrid cases \cite{ref6}.
The discussion will be at the level of the scalar potential, but 
we shall have SUSY in the back of our mind and so occasionally 
refer to the superpotential from which the  potential may be derived. 
In addition we shall assume that the $\phi$ mass arises from
a soft SUSY breaking effect and so has a value of around 1 TeV,
which effectively fixes the scale of the $\psi$ VEV to be around 
$10^{11}$ GeV, as we shall discuss.
In section 3 we shall determine 
exactly how small the dimensionless couplings of inverted hybrid
inflation must be. In section 4 we shall discuss the problems
with obtaining an inverted hybrid potential from a superpotential.

\section{Hybrid and Inverted Hybrid Inflation}

\ind We shall take the potential of hybrid inflation to be of the form
\cite{ref6}:

\begin{eqnarray}
V_\mathrm{H} (\phi,\psi)&=&\frac{1}{4}\lambda_\psi(\psi^2-M^2)^2 +
\frac{1}{2}m_\phi^2\phi^2 + 
\frac{1}{2}\lambda\phi^2\psi^2                              \label{V1h}\\ 
&=&V_0 - \frac{1}{2}m_\psi^2\psi^2 + \frac{1}{4}\lambda_\psi\psi^4 +
\frac{1}{2}m_\phi^2\phi^2 + 
\frac{1}{2}\lambda\phi^2\psi^2, 
\end{eqnarray}

\noind where $\phi$ and $\psi$ are real scalar fields,
$V_0=\frac{1}{4}\lambda_\psi M^4$, $m_\psi^2=\lambda_\psi M^2$ 
and the subscript H stands for ``Hybrid''. This
scalar potential can easily be realized in SUSY. It may be derived
from the superpotential $W=\sqrt{\lambda_\psi/2\;}(M^2-2\Psi^2)\Phi$,
where $\Phi$ and $\Psi$ are complex chiral superfields,
which is the simplest superpotential that breaks a U(1)
symmetry, \cite{ref4}. Adding a soft SUSY-breaking $\phi$ mass term whose mass,
$m_\phi\sim 1$TeV, we
obtain the potential (\ref{V1h}) in terms of the real scalar fields
$\phi$ and $\psi$.

Perhaps the simplest way of achieving inverted hybrid inflation is, as
suggested \cite{ref5,ref6}, 
to just reverse certain signs in this potential, giving

\begin{eqnarray}
V_\mathrm{IH} (\phi,\psi)&=&\frac{1}{4}\lambda_\psi(\psi^2+M^2)^2 -
\frac{1}{2}m_\phi^2\phi^2 - 
\frac{1}{2}\lambda\phi^2\psi^2 + \frac{1}{4}\lambda_\phi\phi^4  \label{V1ih}\\ 
&=&V_0 + \frac{1}{2}m_\psi^2\psi^2 + \frac{1}{4}\lambda_\psi\psi^4 -
\frac{1}{2}m_\phi^2\phi^2 - 
\frac{1}{2}\lambda\phi^2\psi^2 + \frac{1}{4}\lambda_\phi\phi^4,   \label{V2ih}
\end{eqnarray}\\
where we have also added a $\phi^4$ term, enabling
$\phi$ to possess a vacuum expectation value (VEV),
and the subscript IH stands for Inverted Hybrid. 
The necessity of having this term is the essential difference
between the two models. This is a general renormalizable
potential leading to inverted hybrid inflation.

The scalar field $\phi$ in these models acts as the inflaton which slowly
rolls during inflation while the potential is dominated by the vacuum
energy of the $\psi$ field. The field $\psi$ is held in a false 
vacuum $\psi=0$
by the presence of the $\phi$ field, until $\phi$ reaches some
critical value, $\phi_c$ when the effective $\psi$ mass squared becomes
negative, allowing $\psi$ to roll out to its VEV through a second
order phase transition, and inflation consequently ends promptly.

In this case, it is easy to see that for both models,
\begin{equation}
\phi_c=\frac{m_{\psi}}{\sqrt{\lambda}}.        \label{phic}
\end{equation}
In the original hybrid model, $\phi$ starts out greater than this
value and for $\phi>\phi_c$, we shall set $\psi=0$. In this regime,
inflation occurs with the quadratic potential

\begin{equation}
V_\mathrm{H} = V_0+\frac{1}{2}m_\phi^2\phi^2.
\end{equation}

Then, if $V_0$ dominates, inflation ends when $\phi$ becomes less than
$\phi_c$, and the fields reach their true vacuum values; 
$\langle\psi\rangle=M, \langle\phi\rangle=0$, at which the
potential vanishes, and the previous $\psi \leftrightarrow -\psi$
symmetry of the potential is broken.

In the inverted case, $\phi$ is originally less than $\phi_c$, so that
inflation occurs (and the false vacuum exists) when
$\phi<\phi_c$. However now, whilst $\psi=0$ during inflation, the
potential is given by
\begin{equation}
V_\mathrm{IH} = V_0-\frac{1}{2}m_\phi^2\phi^2 +
\frac{1}{4}\lambda_\phi \phi^4,                                 \label{gg}
\end{equation}

\noind which is minimised at
\begin{equation}
\phi_m=\frac{m_\phi}{\sqrt{\lambda_\phi}}.
\end{equation}
Clearly for the hybrid mechanism to work we require 
\begin{equation}
\phi_c < \phi_m,                                                \label{a}
\end{equation}
otherwise $\phi$ would reach its minimum, $\phi_m$ with $\psi$ still
trapped in its false vacuum state, $\psi$ would never reach its VEV
and the flatness conditions (see below) would have been violated long
ago, ending inflation in the usual manner and there would be no
difference to ordinary single-field slow-rollover inflation. If we
impose the stronger condition:
\begin{equation}
\phi_c \ll \phi_m,                                              \label{f}
\end{equation}
then we may neglect the $\phi^4$ term in (\ref{gg}) and achieve the inverted
quadratic potential during inflation;
\begin{equation}
V_\mathrm{IH} = V_0-\frac{1}{2}m_\phi^2\phi^2.                   \label{g}
\end{equation}

\noind From here on we may analyse the two models simultaneously.

The spectral index, $n$, in terms of the slow-roll flatness parameters,
$\epsilon$ and $\eta$ evaluated when cosmological scales leave the
horizon $N$ e-folds before the end of inflation, is given, to lowest order, by
\begin{equation}
(n-1) = -6\epsilon + 2\eta, 
\end{equation}

\noind where 
\begin{equation}
\epsilon \equiv \frac{M_P^2}{16\pi}\left(\frac{V'}{V}\right)^2;
\hspace{0.2in}
\eta \equiv \frac{M_P^2}{8\pi}\left(\frac{V''}{V}\right),
\end{equation}

\noind and for slow-rollover to be a valid approximation, the flatness
conditions, $\epsilon\ll 1, |\eta|\ll 1$ are to be satisfied.

In our models,
\begin{equation}
\epsilon=\frac{1}{16\pi}\frac{M_P^2m_{\phi}^4\phi_N^2}{V_0^2},
\end{equation}
\begin{equation}
|\eta|=\frac{M_P^2}{8\pi}\frac{m_{\phi}^2}{V_0},                   \label{eta}
\end{equation}

\noind where $\phi_N$ is the slow-roll value of the field $N$ e-folds
before the end of inflation, and
$|\eta|=\eta_{\;\mathrm{H}}=-\eta_{\;\mathrm{IH}}$. 

Assuming that 
$V_0$ dominates the potential during inflation, $\phi_N$ is given by 
\begin{equation}
\phi_N = \phi_c e^{\eta N},
\end{equation}

\noind obtained using $N=8\pi M_P^{-2}\int_{\phi_c}^{\phi_N} V/V'\;d\phi$. 
Although $N$ is a
function of the inflationary scale, for all calculational purposes,
this variation is insignificant and we may treat $N$ as constant
taking a value between 40 and 60.

\noind From these results we obtain;
\begin{equation}
\frac{\epsilon }{|\eta |} <\frac{m_{\phi}^2\phi_c^2}{2V_0}\ll 1,
\end{equation}

\noind since $V_0$ dominates the vacuum energy. Thus $\epsilon$ is
negligible compared to $|\eta|$ and we may write,
\begin{equation}
(n-1)_{\mathrm{H}}\;=\;(1-n)_{\mathrm{IH}}\; \approx \;2|\eta|
=\frac{M_P^2}{4\pi}\frac{m_\phi^2}{V_0}.
\end{equation}

\noind Further, the relative contribution of gravitational waves to the CMB
anisotropy given by $R\simeq 12\epsilon$ is therefore negligible. So we are
concerned only with the scalar perturbations and the COBE
normalisation can be written;

\begin{equation}
\left(\frac{\Delta T}{T}\right)_Q^2 = \frac{32\pi}{45}\frac{V^3}{V^{'2}M_P^6},
\end{equation}

\noind where the rhs is to be evaluated $N$ e-folds before the end of inflation
(in this model triggered by $\phi$ reaching $\phi_c$).

\noind The COBE constraint becomes

\begin{equation}
\left(\frac{\Delta T}{T}\right)_Q^2 =
\frac{32\pi}{45}\frac{V_0^3}{m_\phi^4M_P^6\phi_c^2}e^{-2\eta N},
\end{equation}

\noind which is valid for both
hybrid and inverted hybrid cases providing we remember that the
sign of $\eta$ changes between the two cases.

To obtain an order of magnitude estimate of the
problem let us take $\phi \sim
\phi_N \sim \phi_c$. Then as in ref. \cite{ref4} we find:
\begin{equation}
\left(\frac{M}{5.5\times
10^{11}\mathrm{GeV}}\right)
=(\lambda)^{-\frac{1}{10}}(\lambda_\psi)^{-\frac{1}{5}}
\left(\frac{m_\phi}{1\mathrm {TeV}}\right)^{2/5}.
\end{equation}

\noind Thus in SUSY inspired models with $m_{\phi} \sim 1$ TeV
the typical scale $M$ must be of the order $10^{11}$GeV,
and assuming all dimensionless parameters to be of order unity
we see from Eq.\ref{phic} that we also have $\phi_c \sim 10^{11}$ GeV.
Now in hybrid inflation this presents no particular problem,
but in inverted hybrid inflation we see that
$\phi_m \sim 1$ TeV and the condition $\phi_c < \phi_m$ is clearly violated,
as discussed in the introduction. The only way to satisfy this condition
is if we relax our assumption that the dimensionless couplings are of order
unity, as discussed in the next section.

\vspace{0.75in}

\section{The Unnaturally Small Parameters of Inverted Hybrid Inflation}

\ind In this section we shall determine how small the dimensionless
parameters of inverted hybrid inflation must be in order 
to satisfy the condition
$\phi_c < \phi_m$.
To do this we shall keep $V_0$ numerically fixed and apply the COBE
constraint more carefully. The COBE constraint gives a complicated
relation between $m_\phi$ and $\phi_c$, or equivalently between $\eta$
(given in (\ref{eta})) and $\phi_c$ as,

\begin{equation}
\phi_c = \frac{A}{|\eta|}e^{-\eta N},                           \label{b}
\end{equation}
where
$$
A = \frac{1}{\sqrt{90\pi}\left(\frac{\Delta
T}{T}\right)_Q}\frac{V_0^{1/2}}{M_p}. 
$$

In the ordinary hybrid model, $\phi_c$ can take any value, but in the
inverted case, where $\eta$ is negative and (\ref{b}) can be written
$\phi_c = A|\eta|^{-1}e^{|\eta| N}$, there is a minimum value of $\phi_c$ 
corresponding to
\begin{equation}
|\eta| = 1/N,\;\;\; \phi_c = AeN.
\end{equation}
This minimum value corresponds to the most optimistic value possible,
if we are to ensure that $\phi_c < \phi_m$.

From (\ref{phic}) we see that for both potentials, the critical value $\phi_c 
= \sqrt{\frac{\lambda_\psi}{\lambda}}\;M$. In the ordinary hybrid model, we may
proceed as in ref. \cite{ref4} and obtain a curve in parameter space relating
$m_\phi$ and $M$ for a given $\lambda$, $\lambda_\psi$ which require
no fine tuning and may be of order unity.

Unless the inflationary scale, $V_0^{1/4}\geq 10^{13}$GeV, the
spectral index is indistinguishable from 1, without us fine tuning the
coupling constants, $\lambda, \lambda_\psi$. For small $|\eta|$'s, the
contribution to the spectral index from the exponential is negligable
and we may write;

\begin{equation}
(n-1)_{\mathrm{H}}\; \simeq\; 2|\eta|\; \simeq\; \frac{2A}{\phi_c} \sim
10^{4}\frac{V_0^{1/4}}{M_P}\sqrt{\frac{\lambda}{\lambda_\psi^{1/2}}}. 
\end{equation}

\noind For example, if $V_0^{1/4} \sim 10^{11}$GeV, the intermediate
scale, to acquire a significantly
tilted spectrum relating to a spectral index given by $n-1 \sim 0.01$
(the experimental accuracy expected to be obtained in the near
future), we would require $\lambda_\psi \sim 10^{-8} \lambda^2$. 

Returning to our inverted model, although we have an extra free
parameter here, $\lambda_\phi$, we also have equation (\ref{a}) as another
constraint. However, for our convenience, in order that we can neglect the
quartic term and simplify our calculations, we use the stonger
condition $\phi_c \ll \phi_m$ (although the result would still follow
from the weaker condition of equation (\ref{a})). This gives the restriction

\begin{equation}
\lambda_\phi \ll \left(\frac{m_\phi}{\phi_c}\right)^2.             \label{r}
\end{equation}
Using equation (\ref{b}), we may write,

\begin{equation}
\frac{m_\phi}{\phi_c} = B|\eta|^{3/2}e^{-|\eta|N},
\end{equation}
where $B=4\pi\sqrt{45}\left(\frac{\Delta T}{T}\right)_Q$.

Again, holding $V_0$ numerically fixed, and differentiating this with
respect to $|\eta|$, we find the maximum value of 
$m_\phi/\phi_c$ corresponds to; 
\begin{equation}
|\eta|=\frac{3}{2N},
\end{equation}
\begin{equation}
\left(\frac{m_\phi}{\phi_c}\right)_{max}
\simeq 35\;N^{-3/2}\left(\frac{\Delta T}{T}\right)_Q.
\end{equation}
 
\noind Hence we have that: 
\begin{equation}
\lambda_{\phi}\ll\left(\frac{m_\phi}{\phi_c}\right)^2\leq 1.2\times
10^3\; N^{-3}\left(\frac{\Delta T}{T}\right)^2 _Q,
\end{equation}

\noind where the rhs is given by: 
\begin{eqnarray}
rhs & \simeq & 2\times 10^{-12}\hspace{0.3in} \mathrm{using}\;\; 
N=40,\;\left(\frac{\Delta T}{T}\
\right)_Q=10^{-5}\\
rhs & \simeq & 5.6\times 10^{-13}\hspace{.64in} N=60
\end{eqnarray}
Therefore $\lambda_{\phi}$ must be very small indeed in this model.
\vspace{0.25in}

We will now proceed to show that the other coupling constants of the
potential (\ref{V2ih}), $\lambda$ and $\lambda_\psi$ are necessarily
even smaller than $\lambda_\phi$.

The global minimum of the potential (\ref{V2ih}) (at which $V$ does not vanish)
corresponds to;
\begin{equation}
\langle \psi \rangle ^2 =
\frac{\phi_m^2-\phi_c^2}{(\frac{\lambda_\psi}{\lambda}
-\frac{\lambda}{\lambda_\phi})},
\hspace{0.2in}
\langle \phi \rangle ^2 = \frac{\frac{\lambda_\psi}{\lambda}\phi_m^2
-\frac{\lambda}{\lambda_\phi}\phi_c^2}{(\frac{\lambda_\psi}{\lambda}
-\frac{\lambda}{\lambda_\phi})}.
\end{equation}

\noind Taking $\;\langle \psi \rangle ^2>0 \; \Rightarrow\;
\lambda_\psi/\lambda >\lambda/\lambda_\phi$. $\;$ Also, as $\;\langle 
\psi \rangle^2 >\langle \phi \rangle ^2$, we have that   

\begin{equation}
\langle \psi \rangle ^2 \:>\:
\frac{\frac{\lambda_\psi}{\lambda}\phi_m^2
-\frac{\lambda}{\lambda_\phi}\phi_c^2}{(\frac{\lambda_\psi}{\lambda}
-\frac{\lambda}{\lambda_\phi})}
\:>\:
\frac{\frac{\lambda_\psi}{\lambda}(\phi_m^2-\phi_c^2)}
{(\frac{\lambda_\psi}{\lambda}
-\frac{\lambda}{\lambda_\phi})}
= \frac{\lambda_\psi}{\lambda}\langle \psi \rangle^2,
\end{equation}

\noind and thus $\;\lambda_\psi /\lambda < 1$, from which follows
$\;\lambda/\lambda_\phi < 1$.$\;\;$ i.e. $\;\lambda_\phi>
\lambda > \lambda_\psi$. 

The above results have been obtained at tree-level. If radiative corrections
are included then, without SUSY, further fine-tuning will be required
in order to stabilise the potential 
(this is just the usual hierarchy problem.)
Within the framework of a supersymmetric model, radiative corrections to a
hybrid inflation potential have been considered in ref.\cite{ref7},
and a similar analysis would be expected to apply to the inverted
hybrid case. However in the inverted hybrid case, it is more difficult
to obtain the potential from a superpotential, as discussed in the next
section.

\hspace{0.75in}

\section{Inverted Hybrid Superpotentials}

\ind So far we have studied a particular renormalisable potential giving
inverted hybrid inflation and found it not to be viable due to its
unacceptably small parameters. In this section we discuss some of the 
difficulties associated with obtaining a consistent inverted hybrid model
from a superpotential. We have already mentioned the problem
that the $\phi^4$ coupling does not fit in with the idea of flat
directions, so we shall consider a different approach in which
such a coupling is not required.
Our starting point is the superpotential introduced in ref.\cite{ref5}:

\begin{equation}
W=\left(\Lambda^2+\frac{\lambda\Phi^2\Psi^2}{\Lambda^2}\right)\Xi,
\end{equation}

\noind where $\Phi, \Psi$ and $\Xi$ are three complex chiral superfields. The
globally supersymmetric scalar potential, writing
$\Phi=\phi/\sqrt{2},\; \Psi=\psi/\sqrt{2}$, minimising the potential
and assuming $\Xi$ is held at zero, is

\begin{equation}
V=\left(\Lambda^2-\frac{\lambda\phi^2\psi^2}{4\Lambda^2}\right)^2
-\frac{1}{2}m_\phi^2\phi^2+\frac{1}{2}m_\psi^2\psi^2,
\label{pot}
\end{equation}

\noind where $\phi$ and $\psi$ are now real scalar fields and this
form breaks down when
$\lambda\phi^2\psi^2/4\Lambda^2\sim\Lambda^2$. We have also added soft 
SUSY-breaking masses of judiciously chosen signs. 
\footnote{The negative mass squared at low energies 
could be obtained from renormalisation group running of
mass squared which is positive at high energy, for example.}
This potential may be written as:

\begin{equation}
V=V_0-\frac{1}{2}\lambda\phi^2\psi^2
+\frac{\lambda^2\phi^4\psi^4}{16\Lambda^4}
-\frac{1}{2}m_\phi^2\phi^2+\frac{1}{2}m_\psi^2\psi^2.
\end{equation}

\noind where, unlike the previous case, now both 
$m_\psi^2$ and $m_\phi^2$ are of order 1 TeV,
and $V_0=\Lambda^4$.
The inverted mechanism works as before with the same 
critical field value as in Eq.(\ref{phic}).
However in this case there is no minimum $\phi_m$ at $\psi=0$ and
consequently, the restriction (\ref{r}) and its subsequent inference
no longer applies.

On minimising the potential, we find the following conditions
which the fields are supposed to satisfy at their VEVs:

\begin{equation}
\phi^2=4\Lambda^4\frac{(m_\phi^2+\lambda\psi^2)}{\lambda^2\psi^4},
\hspace{0.4in}
\psi^2=4\Lambda^4\frac{(\lambda\phi^2-m_\psi^2)}{\lambda^2\phi^4}
\hspace{0.1in} 
\mathrm{with}\hspace{0.1in} \phi^2>\phi_c^2,
\end{equation}

\noind From
these equations we see that
\begin{equation}
\langle\psi\rangle^2=-\frac{m_\phi^2}{m_\psi^2}\langle\phi\rangle^2,
\end{equation}

\noind which shows that all is not well since if $\langle\phi\rangle^2$
is positive then it follows that
$\langle\psi\rangle^2$ must be negative, which is physically disallowed.

We also find:
\begin{equation}
m_\phi^2\langle\phi\rangle^6
+\left(\frac{4\Lambda^4m_\psi^2}{\lambda}\right)\langle\phi\rangle^2
-\left(\frac{4\Lambda^4m_\psi^4}{\lambda^2}\right)=0,
                                                                   \label{VEV1}
\end{equation}
\begin{equation}
m_\psi^2\langle\psi\rangle^6+\left(\frac{4\Lambda^4m_\phi^2}
{\lambda}\right)\langle\psi\rangle^2
+\left(\frac{4\Lambda^4m_\phi^4}{\lambda^2}\right)=0.
                                                                   \label{VEV2}
\end{equation}
Regarding these equations as cubic in the square of the VEV, it is
trivial to show that while Eq.(\ref{VEV1}) has
a solution with $\langle\phi\rangle^2<\phi_c^2$,
Eq.(\ref{VEV2}) does {\em not} have a
real solution for $\langle\psi\rangle$. 

In fact the potential of 
Eq.(\ref{pot}) is clearly unbounded from below corresponding to
$\frac{\lambda\phi^2\psi^2}{4\Lambda^2}=\Lambda^2$
with $\phi^2 \rightarrow \infty$.
Thus we conclude that this is a very sick model.

\section{Conclusion}

\ind We have argued that while hybrid inflation works perfectly well and fits
in with the idea of flat directions in SUSY, inverted hybrid inflation
faces severe challenges. We have shown that, when inverted hybrid inflation
is expressed in the most straightforward way, as a simple
modification of the hybrid inflation potential, the constraint that
$\phi_c <\phi_m$ can only be satisfied if the dimensionless couplings
are tuned to be very small indeed, say $\lsim 10^{-12}$.
A supersymmetric model which has been proposed and which does
not involve the $\phi^4$ term leads to a potential which is unbounded
from below. We stress that we have not presented a no-go theorem for
models of inverted hybrid inflation, but merely have explored a 
represententative set of such models which have been previously proposed,
and found them to have the stated problems of which the proponents
of these models were unaware. The main point of this paper is to 
expose the challenges faced by such inverted
hybrid inflation models. 
By contrast we find that the original hybrid inflation
scenario is trouble free, and looks very promising.

\begin{center}
{\bf Acknowledgements}
\end{center}
We would like to thank David Lyth for reading a preliminary version
of this manuscript. JS wishes to acknowledge the support of PPARC.

--------------------------------------------------------------------------

\end{document}